\documentclass[a4paper,twocolumn,10pt]{IEEEtran}
\usepackage{}

\usepackage{mathrsfs}

\usepackage{cite}

\usepackage{graphicx}

\usepackage{algorithm}

\usepackage{algorithmic}

\usepackage{psfrag}

\usepackage{amsthm,amssymb}

\usepackage{color}

\usepackage{subfigure}

\usepackage{url}

\usepackage{stfloats}

\usepackage{amsmath}

\usepackage{chngpage}

\usepackage{array}

\newcommand{\picspace}{{\vspace{-0.1 in}}}
\newcommand{\picspacec}{{\vspace{-0.1 in}}}

\usepackage{soul}

\begin{document}

\title{Remote Antenna Unit Selection Assisted Seamless Handover for High-Speed Railway Communications with Distributed Antennas}

\author{Yang Lu$^{\dagger,^\sharp}$, Ke Xiong$^{\dagger,^\sharp}$, Zhuyan Zhao$^{\ddag}$, Pingyi Fan$^{\ast}$, Zhangdui Zhong$^{\dagger}$,
\\
\small
$^\dag$School of Computer and Information Technology, Beijing Jiaotong University, Beijing, P.R. China\\
$^\sharp$National Mobile Communications
Research Laboratory, Southeast University, Nanjing, P.R. China\\
$^\ast$Department of Electronic Engineering, Tsinghua University, Beijing, P.R. China\\
$^\ddag$Nokia Solutions and Networks, P.R. China\\
Corresponding email: kxiong@bjtu.edu.cn}

\maketitle

\begin{abstract}
To attain seamless handover and reduce the handover failure probability for high-speed railway (HSR) communication systems, this paper proposes a remote antenna unit (RAU) selection assisted handover scheme where two  antennas are installed on high speed train (HST) and distributed antenna system (DAS) cell architecture on ground is adopted. The RAU selection is used to provide high quality received signals for trains moving in DAS cells and the two HST antennas are employed on trains to realize seamless handover. Moreover, to efficiently evaluate the system performance, a new metric termed as handover occurrence probability is defined for describing  the relation between handover occurrence position and handover failure probability. We then analyze the received signal strength, the handover trigger probability, the handover occurrence probability, the handover failure probability and the communication interruption probability. Numerical results are provided to compare our proposed scheme with the current existing  ones. It is shown that our proposed scheme achieves better performances in terms of handover failure probability and communication interruption probability.
\end{abstract}


\section{Introduction}

Recently, high-speed railway (HSR) system has developed very fast owing to its convenience and safety. Meanwhile, the wireless service demand during the trip of high-speed train (HST) grows dramatically but this demand is hard to be satisfied, since there are various challenges in the broadband wireless communication for HSR systems\textcolor[rgb]{0,0,0}{\cite{HSR}}. One of the biggest challenges is how to provide seamless handover with low handover failure probability for HSR system.

Firstly, due to the high mobility speed of HST, the handover occurs very frequently which interrupts the wireless connection intermittently. Secondly, the remote base station (BS) providing radio signals with poor strength causes a high handover failure probability and degrades the quality of service, especially, in the boundary region of two neighboring cells.

Aiming at releasing these problems, a lot of research work have been done in the literature\textcolor[rgb]{0,0,0}{\cite{Seamless} -\cite{DualB}}. To release the first problem, two-HST antennas were adopted in\textcolor[rgb]{0,0,0}{\cite{Seamless}}, in which during the handover, the front antenna of the train carries out handover while the rear antenna keeps connecting with BS. After the front antenna completes establishing connection with the target BS, the rear one switches to the working frequency of the target BS. However, due to the near-far signal prorogation effect, the seamless handover of such a configuration is still difficult to be ensured, especially when the base stations are relatively far way from the train. To release the second problem, the distributed antenna system (DAS) cell architecture was introduced to HSR systems\textcolor[rgb]{0,0,0}{\cite{RAU2},\cite{RAU}}, which also improves the system supported data rate as the remote antenna unit deduces the distance between the BS and the train. To inherit the advantages of both two-HST antennas and DAS cells,\textcolor[rgb]{0,0,0}{\cite{DualA}} and\textcolor[rgb]{0,0,0}{\cite{DualB}} investigated them in a single  HSR system to improve the handover performance, where however, only traditional/blanket transmission scheme was considered.

In the traditional/blanket transmission scheme, each BS allocates its available power to its RAUs equally, even during the handover procedure, which neglects the difference of the links between the RAUs and the train antennas. Since the RAUs are usually deployed along the track, there is always one RAU being the closest to the train and other ones are relatively far away from the train. Therefore, the closest RAU should consume all power to achieve a better system performance, rather than all RAUs equally dividing up the power. From this motivation, different from\textcolor[rgb]{0,0,0}{\cite{DualA}} and\textcolor[rgb]{0,0,0}{\cite{DualB}}, we adopt the remote antenna unit (RAU) selection to enhance the handover performance of HSR system based on two-HST antennas and DAS cells.
In our proposed scheme, the available power of each BS is always allocated to the RAU with the minimum pathloss according to the position of the train. As a result, the received signal strength (RSS) at the train can be improved, so does the handover performance.

The contribution of our work can be summarized as follows. \textit{Firstly}, we investigate the handover scheme for HST communication systems,
 which is the first work on introducing the RAU selection in DAS cells to enhance the handover performance of HSR system with two-HST antennas.
  \textit{Secondly}, we present a new performance metric named as handover occurrence probability to describe the relation between handover occurrence
  position and handover performance. Based on this metric, we obtain one significant insight. \textcolor[rgb]{0.00,0.00,0.00}{That is, the appropriate
  handover position can reduce handover failure probability greatly.} \textit{Thirdly}, The system performance of our proposed scheme in handover trigger
  probability\footnote{In other papers, this probability is usually called handover probability.}, handover failure probability and communication interruption
   probability are analyzed in theory and numerical results shows its advantages over the current exiting ones.

The rest of this paper is organized as follows. Section II describes the system model. Section III presents our proposed handover scheme. The analytical results and the numerical results are presented are provided in Section IV and Section V, respectively. Finally, Section VI summarizes this paper.

\section{System Model}

\subsection{System Architecture}
Consider a HSR system, where a train is running at a constant speed $v$ on a straight railway track. The passengers in the train expect a high quality wireless communication service during the trip. To satisfy this requirement and provide broadband wireless communication for the HST, two configurations, (i) two-HST antennas and (ii) DAS cells, are involved in our system, which are described as follows.

\subsubsection{Two-HST antennas} In this configuration, two antennas are deployed on the top of the train which are characterized  by \textit{the front antenna} and \textit{the rear antenna}. The distance of two antennas is set as the length of HST. These two antennas are connected with a Train relay station (TRS) inside the train carriage also by optical fiber links, which is used to provide a cooperative communication service between users and cells along the railway. It means that all information transmitted from (or sent to) the users are aggregated and forwarded by the TRS. By dosing so, all users in the train can be considered as a single ``big user'' connecting with TRS, so that the group handover (i.e., all active users simultaneously requiring handover) from the serving cell to the target cell can be released by executing the handover of the ``big user''. When the train moves within the coverage of one DAS cell, both HST antennas communicate with the serving cells and TRS receives the signals from two antennas by using maximal-ratio combining (MRC) method. When the train moves into the overlap area of two neighboring cells, the front antenna executes handover to the target cell (also termed as the next cell) while the rear antenna keeps connecting with the serving cell (also termed as the current cell), so that a seamless communication service can be provided during handover procedure. Actually, two-HST antennas can not only improve the channel capacity by using space diversity but also help to avoid the penetration loss by using the two-hop BS-TRS-user link transmission.

\subsubsection{DAS cells} 
\begin{figure}
\centering
\includegraphics[width=0.45\textwidth]{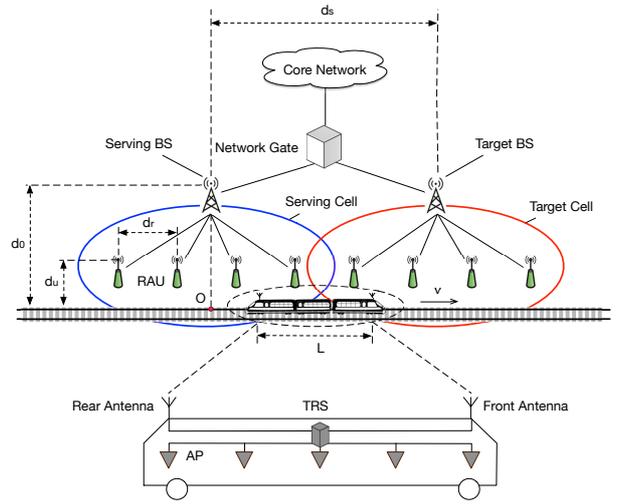}
\caption{System architecture: two-HST antennas and DAS cells.}
\label{system}\picspace
\end{figure}

Different from the traditional cell which has a centralized antenna array at the base station (BS), one DAS cell is formed by a master BS and $N$ RAUs which physically connect with the master BS by optical fiber links. As shown in Fig. \ref{system}, RAUs are distributed along the railway to provide a high-quality seamless coverage of radio signals to HST. The master BS deliver data from core network to each RAU via optical fibers. It is assumed that the optical fibre links are able to provide super high data rate, so the transmission delay between the BS and the RAU is neglected. With each DAS cell, the quality of radio signals over both downlink and uplink can be greatly enhanced by reducing the transmission distance between the serving cell and the antennas of HST. Since there are usually more data transmission in downlink compared with uplink, this paper mainly focuses on the downlink scenario.

 As shown in Fig. \ref{system}, $d_\textrm{s}$ is the distance between two neighboring (master) BSs, $d_\textrm{r}$ is the distance between two neighboring RAUs in one DAS cell, $d_\textrm{0}$ is the distance between (master) BS and railway, $d_\textrm{u}$ is the distance between RAUs and railway, and $L$ is the length of HST. The train location $x$ is measured by the distance between the front antenna and original point which is the cross point between railway track and its vertical line through the serving (master) BS. For clarity, we use $n$ and $i$ to represent the $n$-th RAU of a DAS cell and the $i$-th antenna of HST respectively, where $0 \le n \le N$, and $i\in \{1,2\}$, where ``1'' represents the front antenna and ``2'' represents the tail antenna.

\subsection{Received Signal Strength (RSS)}

RSS is a vital parameter usually adopted in investigating handover scheme. As previously mentioned, the two antennas perform handover separately.

It was proved in \cite{Seamless} that the handover schemes are not sensitive to small-scale fading and the inter-carrier interference (ICI) and noise component could be averaged out of RSS which has no effect on the final average signal strength.  Therefore, at time $t$, the train location $x$ is $v \cdot t$ and the RSS at the $i$-th antenna from the $n$-th RAU\footnote{We only consider RAUs for transmission in a DAS cell.} in the cell $s$ is
\begin{equation}
R_{n,i}^{\left( s \right)}\left( x \right) = {P_n} - {P_L}\left( d_{n,i}^{\left( s \right)}\left(x\right) \right) - \zeta
\label{AAA}
\end{equation}
where $s \in \left\{ {\textrm{TC},\textrm{SC}} \right\}$. $\textrm{SC}$ and $\textrm{TC}$ represent the serving cell and the target cell, respectively. $P_\textrm{n} $ is the transmit power of $n$-th RAU in $s$-th cell, $P_\textrm{n} \le P_\textrm{t}$, $P_\textrm{t}$ is the constraint power of $s$-th cell, $P_L\left(d \right)=A \cdot d^{-\gamma} $ is the path loss, where $A$ is a constant, $d$ is the distance, $\gamma$ is the path loss exponent. $\zeta$ is used to describe the shadow fading which is a normal random variable and $\zeta \sim \mathcal{CN}(0,(\sigma_{n,i}^{( s )})^2)$.

\section{Handover scheme}

This section firstly presents the handover condition based on the RAU selection. Then a dual-cast transmission scheme from the core network to fully utilize the two-HST antennas. At last the details of handover procedure are presented.

\begin{figure}
\centering
\includegraphics[width=0.48\textwidth]{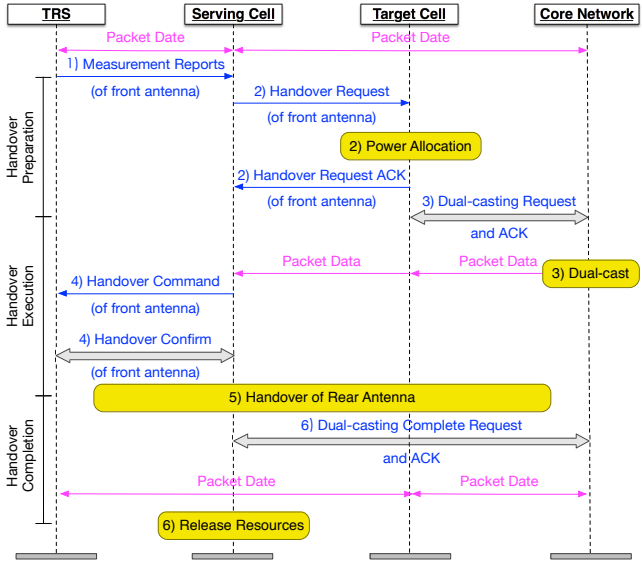}
\caption{Handover procedure of proposed scheme}\picspace
\label{procedure}
\end{figure}
\subsection{Handover Condition}
As there are multiple RAUs in each DAS cell, several transmission schemes, such as the blanket transmission scheme and RAU selection transmission scheme\textcolor[rgb]{0,0,0}{\cite{RAU}}, can be adopted for the information transmission in a DAS cell. The blanket transmission scheme transmits signals through all RAUs and the power is allocated equally for all RAUs. The RAU selection transmission scheme selects one RAU with minimum propagation pathloss for transmission. Compared with blanket transmission scheme, RAU selection transmission scheme is more suitable for HSR system according to\textcolor[rgb]{0,0,0}{\cite{RAU2}}. Thus, we adopt the later one in our handover scheme.

The RAU selection is employed i) for the serving cell to allocate all available power to one RAU for transmission which is closet to the HST;  ii) for the target cell to allocate all power to its first RAU during the handover procedure.

In this case, the RSS at  $i$-th antenna from the $n$-th RAU in $s$-th cell is $R_i^{\left( s \right)}\left(x\right)=\mathop {\max }\limits_i \left\{ {R_{n,i}^{\left( s \right)}\left(x\right)} \right\}$. Since every selected RAU consumes $P_\textrm{t}$ for transmission, the transmit power $P_\textrm{n}$ in (\ref{AAA}) should be modified by $P_\textrm{t}$.

The condition that a handover from the serving cell to the target cell should be triggered by the $i$-th HST antenna is
\begin{equation}
{R_{1,i}^{\left(\textrm{TC}\right)}} - {R_{N,i} ^{\left(\textrm{SC}\right)}}> H
\end{equation}
where ${R_i^{\left(s\right)}}$ is the RSS at the $i$-th antenna from the cell $s$ and $H$ is the protection factor, which is used to avoiding the Ping-Pang effect during handover.

\subsection{Dual-cast}
Since two-HST antennas are employed, during the handover these two antennas shall connect with serving cell and target cell respectively, thus two independent transmission links are constructed.  As for how to keep two links synchronous, we adopt a transmission scheme named as dual-casting which is very similar with the bi-casting\textcolor[rgb]{0,0,0}{\cite{bi-casting}}. In this transmission scheme two copies of data are sent from the core network to both the serving cell and the target cell during handover by optical fiber links which are regarded as no delay. Then the serving cell and the target cell provide the same information to the rear antenna and the front antenna synchronously by RAUs.

\subsection{Handover procedure}
Fig. \ref{procedure} shows the procedures of our proposed handover scheme and the details are given as follows.

\emph{a. Handover preparation}

\begin{enumerate}
\item HST issues the Measurement Report including the RSS of two antennas from TRS to the serving cell periodically. If the serving cell decides to trigger a handover for the front antenna, it sends a handover request message including the selected RAU index to the target cell.
\item The target cell makes an admission control $\&$ configure and allocates transmission power according to the index message. Then it sends a handover request ACK to the serving cell.
\end{enumerate}

\emph{b. Handover Execution}

\begin{enumerate}
\setcounter{enumi}{2}
\item The serving cell continues transmitting data received from the core network to the rear antenna of HST. Meanwhile, the target cell requests one duplicated data from the core network by dual-casting.
\item Once TRS receives the Handover Command of the front antenna from the serving cell, the front antenna breaks the connection with the serving cell and establishes a new connection with the selected RAU in the target cell.
\item When the rear antenna enters the overlap region and receives the handover command from the serving cell which decides to trigger a handover for it based on the Measurement Report, it starts a handover procedure and connects with the target cell.

\end{enumerate}

\emph{c. Handover Completion}

\begin{enumerate}
\setcounter{enumi}{5}
\item When the rear antenna finishes the handover, the serving cell requests to finish dual-casting and the core network makes a request ACK. Then, the data is only transmitted to the target cell and the serving cell releases the resource associated to the TRS.
\end{enumerate}

In this procedure, the handover of two antennas occurs at different time, so it can be ensured that at least one antenna receives data from the core network by the serving cell or target cell. Therefore, the passengers in the train can enjoy an seamless wireless communication service.

\section{Performance Analysis}

To evaluate the performance of the proposed scheme, we analyze four different metrics, i.e., the \textit{handover trigger probability}, the \textit{handover occurrence probability}, the \textit{handover failure probability} and the \textit{communication interruption probability} for it. It is worth noting that in the following analysis, the position of train's front antenna $x$ is regarded as the reference and we consider the period of $\left[ {0,d_\textrm{s}} \right]$.

Before the analysis, we first drive the CDF and the PDF of the transmission link. As mentioned previously, for each cell, only one RAU is selected to transmit signals, so
the CDF of $R_i^{\left( s \right)}\left(x\right)$ can be given by
\begin{flalign}
F_i^{\left( s \right)}\left( x, r \right) &= f_i^{\left( s \right)}\left( {\mathop {\max }\limits_i \left\{ {R_{n,i}^{\left( s \right)}\left(x\right)} \right\} < r} \right) \nonumber\\
&= \prod\nolimits_{n = 1}^N {\Phi \left( {\tfrac{{r - \mu _{n,i}^{\left( s \right)}\left(x\right)}}{\sigma _{n,i}^{\left( s \right)} }} \right)}
\end{flalign}
where $\mu _{n,i}^{\left( s \right)}\left(x\right)=P_t-P_L\left( d_{n,i}^{\left( s \right)}\left(x\right) \right)$ and $\sigma_{n,i}^{\left( s \right)}$ are the mean and standard deviation of $R_i^{\left( s \right)}\left(x\right)$, respectively. Then PDF of $R_i^{\left( s \right)}\left(x\right)$ can be expressed by
\begin{flalign}
f_i&^{\left( s \right)}\left( x,r \right)= \sum\limits_{n = 1}^N {\Phi '\left( {\tfrac{{r - \mu _{n,i}^{\left( s \right)}\left(x\right)}}{\sigma_{n,i}^{\left( s \right)} }} \right)} \mathop \Pi \limits_{j \ne n} \Phi \left( {\tfrac{{r - \mu _{j,i}^{\left( s \right)}\left(x\right)}}{\sigma _{j,i}^{\left( s \right)} }} \right).
\end{flalign}

\subsection{Handover Trigger Probability}
Handover trigger probability is the probability of that the handover trigger condition is met at position $x$. Based on the distribution of RSS and handover condition, the handover trigger probability of $i$-th antenna of HST can be given by (\ref{HTT}).
\begin{flalign}\label{HTT}
&P_i^{\left( \rm{trig} \right)}\left(x,H\right)= \textrm{Pr}\left\{{R_{1,i}^{\left( \textrm{TC} \right)}\left(x\right) -
R_{N,i}^{\left( \textrm{SC} \right)}\left(x\right) > H} \right\}\\
&= \textrm{Pr}\left\{ {R_{1,i}^{\left( \textrm{TC} \right)}\left(x\right) > r + H} \right\}\textrm{Pr}\left\{ {
 R_{N,i}^{\left( \textrm{SC} \right)}\left(x\right) = r} \right\}\nonumber\\
&= \int\nolimits_r {\left( {1 - F_i^{\left( \textrm{SC} \right)}\left(x, {r + H} \right)} \right)f_i^{\left( \textrm{TC} \right)}\left( x, r \right)} dr\nonumber
\end{flalign}

Since the handover is triggered between the last RAU of the serving cell and the first RAU of target cell in RAU selection scheme, $P_i^{\left( \rm{trig} \right)}\left(x,H\right)$ can be rewritten as
\begin{equation}\label{AHO}
P_i^{\left( \rm{trig} \right)}\left(x,H\right) = \textrm{Pr}\left\{ {R_{1,i}^{\left( \textrm{TC} \right)}\left(x\right) - R_{N,i}^{\left( \textrm{SC} \right)}\left(x\right) > H} \right\},
\end{equation}
where $R_{1,i}^{( \textrm{TC} )}(x)\sim \mathcal{CN}(P_t-P_L( {d_{1,i}^{( \textrm{TC} )}(x)} ), (\sigma_{1,i}^{( \textrm{TC} )})^2 )$ and $R_{N,i}^{( \textrm{SC} )}(x)\sim\mathcal{CN}(P_t-P_L( {d_{N,i}^{( \textrm{SC} )}(x)} ), (\sigma_{N,i}^{( \textrm{SC} )})^2 )$. As the sum of Gaussian distribution still follows Gaussian distribution, (\ref{AHO}) can be expressed by
\begin{equation}\label{AHO2}
 P_i^{\left( \rm{trig} \right)}\left(x,H\right) =1 - \Phi \left( {\tfrac{{H - \left( {{P_L}\left( {d_{N,i}^{\left( \textrm{SC} \right)}\left(x\right)} \right) - {P_L}\left( {d_{1,i}^{\left( \textrm{TC} \right)}\left(x\right)} \right)} \right)}}{{\sqrt {{{\left( {\sigma _{N,i}^{\left( \textrm{SC} \right)}} \right)}^2} + {{\left( {\sigma _{1,i}^{\left( \textrm{TC} \right)}} \right)}^2}} }}} \right)\nonumber
\end{equation}

\subsection{Handover Occurrence Probability}

Handover occurrence probability is the probability of that the handover occurs at position $x$, which means that at position $x$ the handover condition is met and before it the handover is not triggered (i.e., the handover condition is not met before $x$). The handover occurrence probability of the $i$-th antennas on HST in the proposed scheme can be given by (\ref{HOTP2}).

\begin{figure*}[!hb]
\picspace
\hrule
\begin{flalign}\label{HOTP2}
P_i^{\left( {\textrm{occ}} \right)}\left( {x,H} \right)&= \textrm{Pr}\left\{ {R_{1,i}^{\left( \textrm{TC} \right)}\left( x \right) - R_{N,i}^{\left( \textrm{SC} \right)}\left( x \right) > H\left| {R_{1,i}^{\left( \textrm{TC} \right)}\left( \tau \right) - R_{N,i}^{\left( \textrm{SC} \right)}\left( \tau \right) > H,\tau < x} \right.} \right\} \\
&= \textrm{Pr}\left\{ {R_{1,i}^{\left( \textrm{TC} \right)}\left( x \right) - R_{N,i}^{\left( \textrm{SC} \right)}\left( x \right) > H} \right\}\textrm{Pr}\left\{ {R_{1,i}^{\left( \textrm{TC} \right)}\left( \tau \right) - R_{N,i}^{\left( \textrm{SC} \right)}\left( \tau \right) > H,\tau < x} \right\} = P_i^{\left( \rm{trig} \right)}\left( {x,H} \right)\int {P_i^{\left( \rm{trig} \right)}\left( {\tau ,H} \right)d\tau }\nonumber
\end{flalign}
\hrule
\picspace
\end{figure*}

\subsection{Handover Failure Probability}
Handover failure occurs when the handover condition is met but the average RSS from the target cell is lower than a threshold $T$ which is the minimum RSS to maintain a wireless communication. So the handover failure probability of the $i$-th antenna of HST can be expressed as
\begin{flalign}
&P_i^{\left( \rm{fail} \right)}\left(x,H,T\right) \\
& = \textrm{Pr}\left\{ {\left. {R_{1,i}^{\left( \textrm{TC} \right)}\left(x\right) < T} \right|R_{1,i}^{\left( \textrm{TC} \right)}\left(x\right) - R_{N,i}^{\left( \textrm{SC} \right)}\left(x\right) > H} \right\}\nonumber\\
 & = \tfrac{{\textrm{Pr}\left\{ {R_{1,i}^{\left( \textrm{TC} \right)}\left(x\right) < T,R_{1,i}^{\left( \textrm{TC} \right)}\left(x\right) - R_{N,i}^{\left( \textrm{SC} \right)}\left(x\right) > H} \right\}}}{{\textrm{Pr}\left\{ {R_{1,i}^{\left( \textrm{TC} \right)}\left(x\right) - R_{N,i}^{\left( \textrm{SC} \right)}\left(x\right) > H} \right\}}}\nonumber\\
 & = \tfrac{{\textrm{Pr}\left\{ {R_{N,i}^{\left( \textrm{SC} \right)}\left(x\right) < r - H\left| {R_{1,i}^{\left( \textrm{TC} \right)}\left(x\right)} \right. = r,R_{1,i}^{\left( \textrm{TC} \right)}\left(x\right) < T} \right\}}}{{P_i^{\left( \rm{trig} \right)}\left(x,H\right)}}\nonumber\\
 & = \tfrac{1}{{P_i^{\left( \rm{trig} \right)}\left(x,H\right)}}\int_{ - \infty }^T {F_i^{\left( \textrm{SC} \right)}\left( x,{r - H} \right)f_i^{\left( \textrm{TC} \right)}} \left(x, r \right)dr\nonumber
\end{flalign}
and it can be rewritten as (\ref{AHF}) for our proposed handover transmission scheme.
\begin{figure*}[!ht]
\picspace
\begin{flalign}\label{AHF}
P_i^{\left( \rm{fail} \right)}\left( {x,H,T} \right) &= \textrm{Pr}\left\{ {\left. {R_{1,i}^{\left( \textrm{TC} \right)}\left( x \right) < T} \right|R_{1,i}^{\left( \textrm{TC} \right)}\left( x \right) - R_{N,i}^{\left( \textrm{SC} \right)}\left( x \right) > H} \right\}= \tfrac{\textrm{Pr}\left\{ {\left. {R_{N,i}^{\left( \textrm{SC} \right)}\left( x \right) < r - H} \right|R_{1,i}^{\left( \textrm{TC} \right)}\left( x \right) = r,R_{1,i}^{\left( \textrm{TC} \right)}\left( x \right) < T} \right\}}{\tilde P_i^{\left( \rm{trig} \right)}(x, H)}\nonumber\\
&= \frac{1}{{P_i^{\left( \rm{trig} \right)}\left( {x,H} \right)}}\int_{ - \infty }^T {Q\left( {\tfrac{{r - H - \left( {{P_t} - {P_L}\left( {d_{N,i}^{\left( \textrm{SC} \right)}\left( x \right)} \right)} \right)}}{{\sigma _{N,i}^{\left( \textrm{SC} \right)}}}} \right)} \frac{1}{{\sqrt {2\pi } \sigma _{1,i}^{\left( \textrm{TC} \right)}}}{\exp\left( - {{\left( {\tfrac{{r - \left( {{P_t} - {P_L}\left( {d_{1,i}^{\left( \textrm{TC} \right)}\left( x \right)} \right)} \right)}}{{\sqrt 2 \sigma _{1,i}^{\left( \textrm{TC} \right)}}}} \right)}^2}\right)}dr
\end{flalign}
\hrule
\end{figure*}

\subsection{Communication Interruption Probability}

Communication interruption of $i$-th antenna means that during the trip of HST, RSS from the serving cell\footnote{When $i$-th antenna finishes the handover, its former target cell becomes the serving cell.} is lower than $T$, resulting in  wireless link failure. It is an important metric to evaluate the quality of wireless communication service and can be given by
\begin{flalign}
P_i^{\left( \rm{out} \right)}\left(x,T\right) &= \textrm{Pr}\left\{ {\mathop {\max }\limits_s \left\{ {R_i^{\left( s \right)}\left(x\right)} \right\} < T} \right\}\\
 &= \mathop {\min }\limits_s \textrm{Pr}\left\{ {R_i^{\left( s \right)}\left(x\right) < T} \right\}= \mathop {\min }\limits_s F_i^{\left( s \right)}\left(x, T \right)\nonumber
\end{flalign}

As for our proposed handover scheme, the communication interruption occurs only when both two antennas fail to connect to the serving cell. As the channels between the two antennas and the BS are independent, the communication interruption probability can be given by
\begin{flalign}\label{POUT}
{P^{\left( \rm{out} \right)}\left(x, T \right)} & = \prod\nolimits_{i = 1}^2 {P_i^{\left( \rm{out} \right)}\left(x, T \right)}\\
& = \prod\nolimits_{i = 1}^2 {\mathop {\min }\nolimits_s F_i^{\left( s \right)}\left(x, T \right)}\nonumber
\end{flalign}
It can be seen from (\ref{POUT}) that two-HST antennas is useful in improving the communication interruption performance during the trip of the HST.

\section{Numerical Results and Discussion}
 Some numerical results are provided to show the performance of our proposed handover scheme here. For comparison, three other schemes are also simulated, i.e., i) two-HST antennas and DAS cells with blanket transmission scheme (DAS-B), ii)  single HST antenna and DAS cells with RAU selection transmission scheme (DAS-S) and iii) two-HST antennas and traditional cells. Particularly, as for the RSS of DAS-B, it is a sum of lognormal variables which does not have a closed-form probability distribution function, so we use the moment generating function (MGF)\textcolor[rgb]{0,0,0}{\cite{MGF}} to approximate it. In the presented simulation results, each point of the curves was averaged over 1000 realizations. The parameters used in simulations are summarized in Table \ref{parameter}.
\begin{table}[!ht]
\centering
\caption{Simulation parameter}
\begin{tabular}{ lc }
\hline
Parameters & Value \\
\hline
Transmit power of each cell $(P_\textrm{T})$ & $86dBm$ \\
Shadow fading deviation $(\sigma)$ & $4dB$ \\
Path loss model $P_L\left(d\right)$ & ${\rm{31}}.{\rm{5 + 35lo}}{{\rm{g}}_{10}}\left( d \right)$ \\
Distance between two cells $(d_\textrm{s})$ & $3000m$ \\
Number of RAUs in one cell $(N)$ & $4$ \\
Noise density $(N_\textrm{0})$ & $145dBm/Hz$ \\
Signal threshold $(T)$ & $-30dB$ \\
Hysteresis margin $(H)$ & $2dB$ \\
Distance from BS to railways ($d_\textrm{0}$) & $100m$ \\
Distance from RAU to railways ($d_\textrm{u}$) & $60m$ \\
Length of train ($L$) & $200m$\\
Speed of train ($v$) & $100m/s$ \\
\hline
\label{parameter}
\end{tabular}
\end{table}

\begin{figure}
\centering
\includegraphics[width=0.42\textwidth]{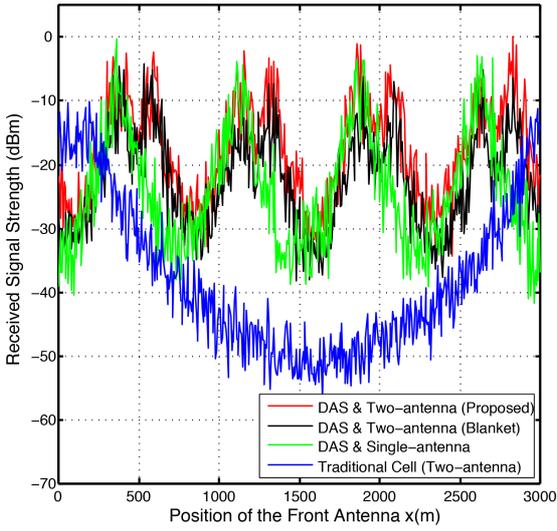}
\caption{RSS versus position of the front antenna $x$. }
\label{RSS}\picspacec
\end{figure}

Fig. \ref{RSS} shows RSS of four different schemes during the period of interest, i.e., $\left[ {0,d_\textrm{s}} \right]$. It can be seen that the proposed scheme has the highest RSS in most time and the DAS-B has the second-highest RSS. Compared with DAS-S, employing two-HST antennas in our proposed scheme yields a higher RSS. Compared with traditional cells which only has a higher RSS for $x=0$ and $x=3000$, the proposed scheme can improve the quality of RSS by reducing the transmission distance with RAUs.

\begin{figure}
\centering
\includegraphics[width=0.42\textwidth]{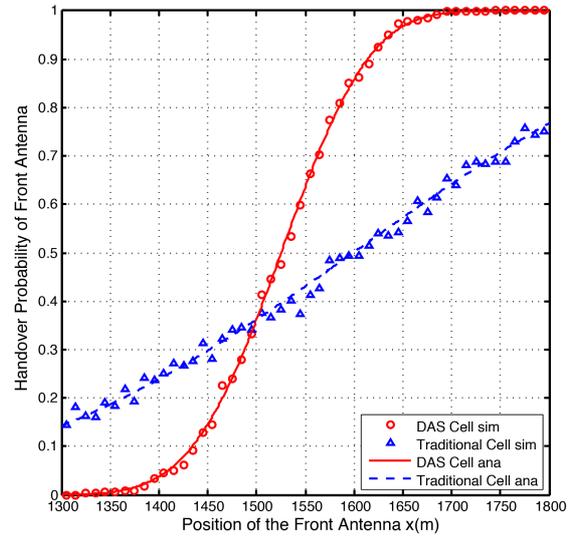}
\caption{Handover trigger probability during the overlap.}
\label{HO}\picspacec
\end{figure}

\begin{figure}
\centering
\includegraphics[width=0.42\textwidth]{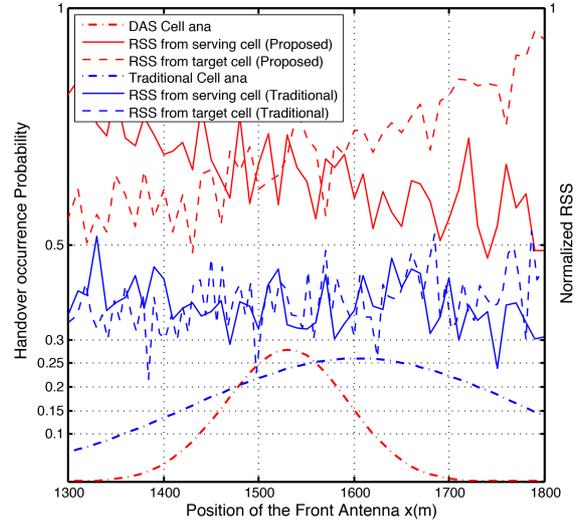}
\caption{Handover occurrence probability during the overlap.}
\label{HOT}\picspacec
\end{figure}

Fig. \ref{HO} and Fig. \ref{HOT} present the handover trigger probability and handover occurrence probability versus the position $x$. It is known that an efficient handover scheme expects handover occurrence position around the middle point of two neighboring cells. Since there is a protection factor, the appropriate handover occurrence point shall shift a little to the target cell. The optimal handover trigger probability should change from $0$ to $1$ dramatically when the train passes the appropriate handover occurrence point. In this sense, our proposed scheme does obtain a better performance than traditional one.
It is noted that all the  curves intersect around the point with x-coordinate of 1500 in Fig. \ref{HO}, which is the middle point of the two neighboring BS.

\begin{figure}
\centering
\includegraphics[width=0.42\textwidth]{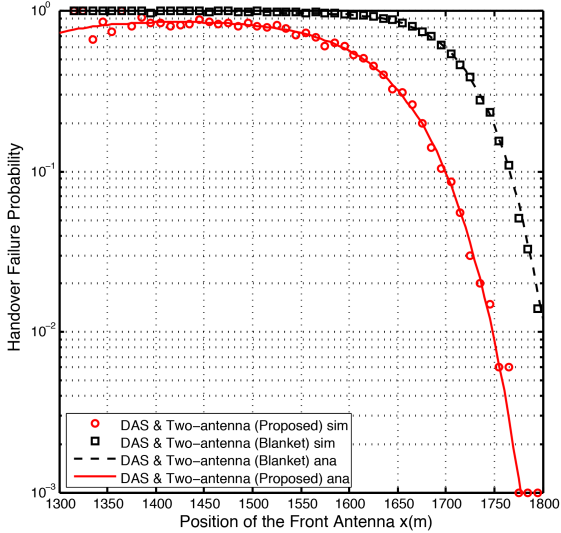}
\caption{Handover failure probability during the overlap.}
\label{HF}\picspacec
\end{figure}

In Fig. \ref{HF}, the proposed scheme achieves a lower handover failure probability compared with DAS-B in the boundary region. Since the handover trigger probability of these two schemes is similar, the difference in this performance is caused by the variance of RSS between target cell and serving cell. The proposed scheme can achieve a better performance because the RAU selection transmission scheme always allocates the power to the RAU closet to HST and makes the transmission suffer minimum pathloss. As for DAS-S and traditional cell, the former one has a similar performance in handover failure probability with the proposed scheme and the later one always suffers a very high handover failure probability since the BSs are too far away from HST when the train is moving in the boundary region. Based on Fig. \ref{HOT} and Fig. \ref{HF}, a very important insight is observed that there exists an appropriate position with high-quality received signals for activating the handover to get very low handover failure probability. Such a position is often near to the middle point of the overlap of two neighboring cells and highly depends on the handover trigger RSS protection factor.

\begin{figure}
\centering
\includegraphics[width=0.42\textwidth]{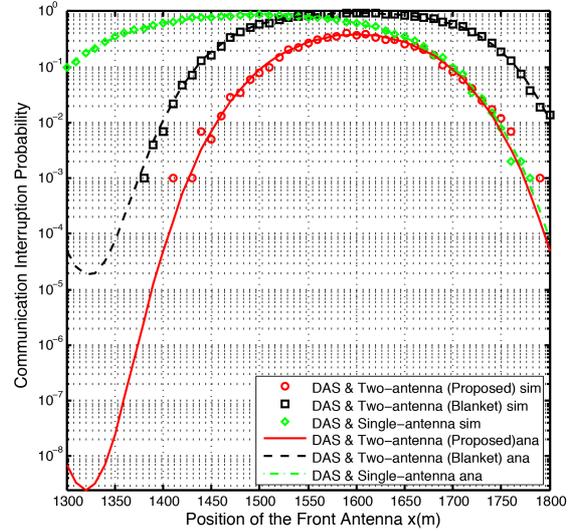}
\caption{Communication Interruption Probability during the overlap.}
\label{HT}\picspacec
\end{figure}

Fig. \ref{HT} plots the communication interruption probability of three different schemes all adopting DAS cells during the overlap. It can be observed that our proposed scheme has the lowest communication interruption probability among the four discussed schemes.  This certificates that the RAU selection transmission scheme provides a better RSS.

\section{Conclusion}
This paper proposed a RAU selection assisted handover scheme for HSR system, which adopted two-HST antennas and DAS cells to provide seamless wireless coverage and improve the quality of RSS so that the handover failure probability could be reduced. The expressions of the received signal strength, the handover trigger probability, the handover occurrence probability, the handover failure probability and the communication interruption probability of our proposed method were derived. Both the analytical and simulation results show that the proposed handover scheme achieves a better handover performance compared with the current existing ones.

\section*{Acknowledgment}
This work was supported by the National
Basic Research Program of China (973 Program), no. 2012CB316100(2), by the Open Research
Fund of National Mobile Communications Research Laboratory, Southeast University, no. 2014D03 and also by the Beijing Natural Science Foundation, no. 4162049.

\end{document}